\documentclass[vecphys]{svmult}

\usepackage{makeidx}         
\usepackage{graphicx}        
\usepackage{multicol}        
\usepackage[bottom]{footmisc}

\makeindex             

\begin{document}

\title*{The ESO Large Programme ``First Stars''}
\authorrunning{Bonifacio et al.}
\author{P. Bonifacio\inst{1,2,3}\and
J. Andersen\inst{4,5}
\and
S.M. Andrievsky\inst{6}
\and
B. Barbuy\inst{7}
\and
T.C.~Beers\inst{8}
\and
E. Caffau\inst{2}
\and
R.~Cayrel\inst{2}
\and
E. Depagne\inst{9}
\and
P. Fran\c cois\inst{2}
\and
J.I.~Gonz\'alez Hern\'andez\inst{1,2}
\and
C.J.~Hansen\inst{10}
\and
F. Herwig\inst{11}
\and
V. Hill\inst{2}
\and
S.A.~Korotin\inst{6}
\and
H.-G.~Ludwig\inst{1,2}
\and
P.~Molaro\inst{3}
\and
B.~Nordstr\"om\inst{4}
\and
B. Plez\inst{12}
\and
F.~Primas\inst{10}
\and
T.~Sivarani\inst{8}
\and
F. Spite\inst{2}
\and
M.~Spite\inst{2}
}
\institute{CIFIST Marie Curie Excellence Team
\and 
GEPI, Observatoire de Paris, CNRS, Universit\'e Paris Diderot; Place
Jules Janssen 92190
Meudon, France
\and
Istituto Nazionale di Astrofisica,
Osservatorio Astronomico di Trieste,  Via Tiepolo 11,
I-34143 Trieste, Italy
\and
The Niels Bohr Institute, Astronomy, Juliane Maries Vej 30,
DK-2100 Copenhagen, Denmark
\and
Nordic Optical Telescope, Apartado 474, E-38700 Santa Cruz de 
La Palma, Spain
\and
Department of Astronomy and 
Astronomical Observatory, 
Odessa National University, Shevchenko Park, 65014 Odessa, Ukraine
\and
Universidade de Sao Paulo, Departamento de Astronomia,
Rua do Matao 1226, 05508-900 Sao Paulo, Brazil
\and
Department of Physics \& Astronomy and JINA: Joint Institute for Nuclear
Astrophysics, Michigan State University,
East Lansing, MI 48824, USA
\and
Las Cumbres Observatory, Santa Barbara, California, USA
\and
European Southern Observatory (ESO),
Karl-Schwarschild-Str. 2, D-85749 Garching b. M\"unchen, 
Germany
\and
Keele Astrophysics Group, School of Physical 
and Geographical Sciences, 
Keele University, Staffordshire ST5 5BG, UK
\and
GRAAL, Universit\'e de Montpellier II, F-34095 
Montpellier
Cedex 05, France
}
\maketitle

\section{Introduction}
\label{sec:1}

In ESO period 65 (April-September 2000) 
the large programme 165.N-0276, led by Roger Cayrel, began
making use of UVES  at the Kueyen VLT telescope.
Known within the Team and outside as ``First Stars'', it
was aimed at obtaining high resolution, high signal-to-noise ratio
spectra in the range 320 nm -- 1000 nm for a large sample
of extremely metal-poor (EMP) stars identified from the HK objective prism
survey \cite{beers85,beers92}.
The goal was to use these spectra to determine accurate
atmospheric parameters and chemical composition of these
stars which are among the oldest objects amenable to our
detailed study.
Although these stars are not the first generation of stars
they must be very close descendants of the first generation.
One may hope to gain insight on the nature of the progenitors
from detailed information on the descendants.

The extremely metal-poor stars are very rare objects and 
finding them in large numbers requires specially designed 
surveys. All of the proponents of the large programme had
been actively working on the medium-resolution follow-up
of the HK survey (results still to be published), from either
ESO La Silla, Kitt Peak or Roque de los Muchachos.
Such a follow-up is mandatory in order to obtain a good list
of candidates on which one can invest the time of an 8 m telescope.

The programme was allocated a total of 39 nights between periods 65
and 68, these were split into 8 observational runs of unequal length.
The observations were carried out in visitor mode because 
UVES was used in non-standard settings. 
The settings selected were Dic1 396+573 and Dic1 396+850,
typically with a $1''$ slit for a resolution $R\sim 43000$.
These settings were preferred over the standard Dic1 390+580
and Dic1 390+860, because you gain the Ba\,{\sc ii} 455.4 nm
line in the blue, Zn\,{\sc i} 471 nm in the red and
Li\,{\sc i} 670.8 nm in the reddest setting.

The main results of the large programme are published on a series
of papers ``First Stars'' on A\&A, so far 10 papers have been published,
one is in press, a few more in preparation. In addition a number
of papers not in the ``series'' have been published, up to know
there are 19 refereed papers published, which make use of the
data acquired in the course of this large programme.

In this contribution we highlight the main results of the large programme.

\section{Uranium in a EMP star}
\label{sec:2}

The first surprise came quite early in the programme,
Vanessa Hill was conducting the observations in August 2000
when she realised, from the quick look data, 
that the giant CS 31082-001 had an exceptional 
spectrum, characterised by extremely low metallicity and a large
enhancement of  the r-process elements. She was in fact
able to identify immediately the Th{\sc ii} 401.9 nm line,
which displayed a remarkable strength. This induced her in the
following nights to acquire blue spectra of higher resolution
with slicer \# 2  in the hope of being able to identify
and perhaps measure uranium in this star.
The star is now often colloquially dubbed as ``Hill's star'',
and in fact her intuition proved correct since this was the
first metal-poor star for which it was possible
to measure the uranium abundance, opening up a new
possibility for nucleochronology\cite{cayrel01,hill}.
This star actually showed how little we knew on the r-process.
While Th/U proved to be a reliable chronometer Eu/Th
and Eu/U provided unrealistically small ages, using
the then available production ratios.
Also Pb in this star is a real puzzle, in fact
the majority of lead in this star is what
you expect from the decay of Th and U, leaving
very little space for Pb production during the
r-process\cite{plez}.

\section{The Spite plateau at the lowest metallicities}
\label{sec:3}

Ever since Monique and Fran\c cois Spite discovered
that warm metal-poor stars share the same Li abundance
(the {\em Spite plateau})\cite{spite82a,spite82b}, there
has been an active research on this field. 
What we wish to understand is if this {\em plateau}
indicates the primordial Li abundance, as 
initially proposed\cite{spite82a,spite82b}, or not.

  \begin{figure}
\centering  \resizebox{\hsize}{!}{\includegraphics[clip=true]{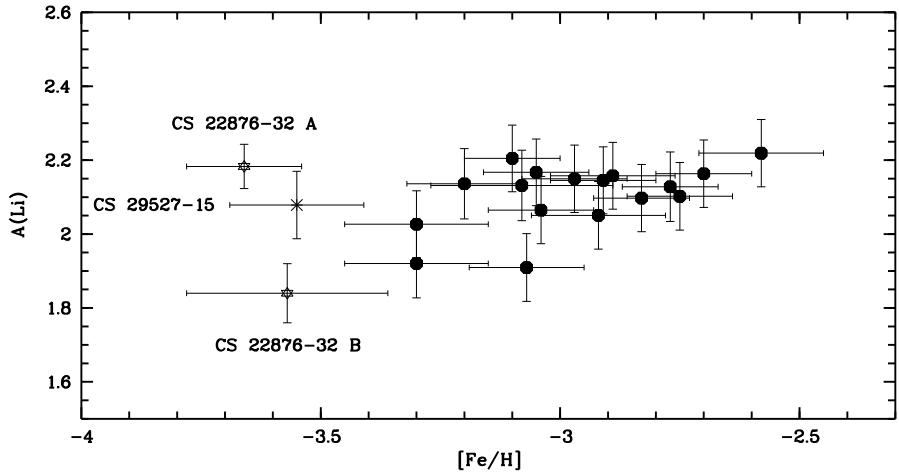}}
\caption{The Spite plateau at the lowest metallicities as portrayed by the
``First Stars'' data. The stars whose names are labelled are binaries, 
for CS 22876-32 an orbital solution is available and the analysis has been
done taking properly into account the veiling and Li in both
components has been measured, for CS22957-15 this has not
been possible, due to the lack of the necessary data, however 
the correction for the veiling is likely not very large.}
 \label{spitepl}
 \end{figure}

The ``First Stars'' large programme allowed to explore 
the {\em Spite plateau} at the lowest known metallicities.
There are no known dwarf stars with a metallicity (meant as
$Z$, total metallicity, not [Fe/H]) lower than the
stars shown in Fig.\ref{spitepl}.
The data are those of \cite{B07} and \cite{jonay}, the picture
which emerges is that the plateau seems to continue
at the lowest metallicities. It is possible that there is a larger
scatter, however the impact on this picture of stars in which
lithium may have been partially depleted is yet unclear.
The difference in lithium content between the two components
of the binary system CS 22876-32 has no clear explanation.
The cooler component (star B) has an effective temperature
of 5900 K and should not display Li depletion according
to standard models.

From the cosmological point of view there is a tension between
the value of the {\em Spite plateau}, A(Li)$\sim 2.1$
and the value predicted by standard big bang nucleosynthesis,
when the baryonic density derived from the power spectrum
of the fluctuations of the cosmic microwave background is
used\cite{spergel,B07}, A(Li)$= 2.64$.
Several ways to explain this discrepancy have been suggested,
and generally they go in two possible directions:
a) the {\em Spite plateau} does not represent the 
primordial abundance or b) primordial nucleosynthesis
did not proceed as assumed in the ``standard'' model.
At present both solutions are possible and
further observations of EMP stars, to understand if
there is an excess scatter of Li at the lowest metallicities,
could give useful indications.

\section{Abundance ratios, what did we learn ?}
\label{sec:4}

When we started the large programme, several of us,
were expecting that at the lowest metallicities
we would begin to see the effects of the pollution
of very few supernovae(SNe), possibly a single supernova.
As a consequence we were expecting considerable
scatter in the abundance ratios, which would be 
the signature of the different masses of the polluting SNe
and incomplete mixing of the gas in the early Galaxy.
To the surprise of several of us we found instead that
the majority of elements C to Zn display a remarkable uniformity,
with well defined trends with metallicity\cite{cayrel04}. The scatter
in these trends can be totally explained by observational
error. One explanation of this low scatter is an efficient
mixing of the early Galaxy.
Alternatively one could argue in favour of a narrow range
of masses of SNe actually contributing to chemical enrichment.


The  exceptions, among lighter elements, were Na and Al, that
displayed a star to star 
scatter larger than observed for other elements.
This excess scatter also made the definition of trends somewhat ambiguous.
A reanalysis of both elements using full NLTE line formation
was in fact able to solve the problem\cite{A07,A08}.
Sodium appears to be constant with metallicity among EMP stars,
with [Na/Fe]$ = −0.21\pm  0.13 $, and the same is true for aluminium 
with [Al/Fe] $= −0.08\pm  0.12$.


The observations of C and N, showed that
in a [C/Fe] vs. [N/Fe] diagram the giants split nicely
into two groups, one with high [N/Fe] and low [C/Fe],
which we call ``mixed'',
the other with lower [N/Fe] and higher [C/Fe
which we call ``unmixed''\cite{spite05,spite06}.
As expected from the theory of stellar
evolution ``mixed' stars are typically the more luminous giants,
although there are a few exceptions.


At variance with lighter elements the n-capture elements
display a large scatter, which cannot be explained 
by observational errors\cite{F07}.
Such scatter, coupled with the very uniform
ratios of the lighter elements demands an inhomogeneous
chemical evolution.
Already from the results of CS 31082-001 it was clear 
that the r-process is not ``universal'' and several r-processes
may be needed.
The data on n-capture elements clearly indicates that a {\em second} r-process
is the main production channel  at [Ba/H]$<-2.5$.


Among the dwarf stars  we found four
which were C enhanced \cite{S04,S06}.
In all of them the C-enhancement has come about 
as a consequence of mass-transfer from an AGB companion.
The abundance pattern of n-capture elements is rather
diverse among the stars, suggesting nucleosynthesis taking place under 
different physical conditions.


The giant CS 22949-037 is one of the most extraordinary found in
the course of the large programme\cite{depagne}.
With [Fe/H]$\sim -4.0$ it is one of the most iron
poor stars found in the sample, however its 
high abundances of CNO ([O/Fe]$\sim +2$, [C/Fe]$\sim +1.2$,
[N/Fe]$\sim +2.6$) make its global metallicity $Z$ not so
extreme as that of the four giants with [Fe/H]$\sim -4$\cite{F03},
which are, so far, the star with the lowest $Z$ known.
There is no totally satisfactory model to explain
the abundance pattern in CS 22949-037, however
it is clear that some special kind of SN is needed to
explain such an extraordinary pattern. 

\section{Needs for the future}

One would like to extend the work done so far, with high resolution,
high S/N ratio spectra of stars of even lower metallicities. Such
stars should be found by on-going and projected surveys (SEGUE, LAMOST, SkyMapper...).
Most of these are however expected to be around 18th magnitude or fainter, UVES
can work at these faint magnitudes, but...slowly.
The proposed high resolution spectrograph ESPRESSO (see Pasquini this meeting)
in the mode combining the 4UTs, would be ideal for these targets.
According to the preliminary estimates ESPRESSO 4UTs, at a resolution
of $R\sim 45000$  should beat UVES in efficiency for all targets
fainter than V=17.5, when read-out-noise begins to be important.



\printindex
\end{document}